# Local inversion-symmetry breaking in a bismuthate high-$T_c$ superconductor

S. Griffitt[1*], M. Spaić[2*], J. Joe[1], Z. Anderson[1], D. Zhai[1], M. J. Krogstad[3], R. Osborn[3], D. Pelc[1,2†], and M. Greven[1†]

[1]School of Physics and Astronomy, University of Minnesota, Minneapolis, MN, USA

[2]Department of Physics, Faculty of Science, University of Zagreb, Zagreb, Croatia

[3]Materials Science Division, Argonne National Laboratory, Lemont, IL, USA

*authors contributed equally

†correspondence to: dpelc@phy.hr, greven@umn.edu

**The doped perovskite BaBiO$_3$ exhibits a maximum superconducting transition temperature ($T_c$) of 34 K and was the first high-$T_c$ oxide to be discovered, yet pivotal questions regarding the nature of both the metallic and superconducting states remain unresolved. Although it is generally thought that superconductivity in the bismuthates is of the conventional $s$-wave type, the pairing mechanism is still debated, with strong electron-phonon coupling and bismuth valence or bond disproportionation possibly playing a role. Here we use diffuse x-ray scattering and Monte Carlo modeling to study the local structure of Ba$_{1-x}$K$_x$BiO$_3$ across its insulator-metal boundary. We find no evidence for either long- or short-range disproportionation, which resolves a major conundrum, as disproportionation and the related polaronic effects are likely not relevant for the metallic and superconducting states. Instead, we uncover nanoscale structural correlations that break inversion symmetry, which has far-reaching implications for the electronic physics, including the pairing mechanism. This unexpected finding furthermore establishes that the bismuthates belong to the broader classes of materials with hidden spin-orbit coupling and a tendency towards inversion-breaking displacements.**

Similar to other perovskite and perovskite-derived superconducting oxides, such as strontium titanate, strontium ruthenate and the high-$T_c$ cuprates, the basic structural unit in the (nearly) cubic bismuthates is a metal-oxygen octahedron (Fig. 1) [1,2]. The parent compound BaBiO$_3$ is an insulator with commensurate charge-density-wave (CDW) order that is thought to be caused by either a periodic modulation of the effective bismuth valence [2] or a Bi-O charge-transfer mechanism with charge localization on the oxygen orbitals [3,4]. With hole doping, this modulated phase quickly disappears, and a superconducting metal appears at sufficiently high hole concentrations. Due to strong electron-lattice coupling, the CDW in the parent compound involves three-dimensional oxygen breathing distortions of the octahedra (Fig. 1d). One of the prevailing hypotheses regarding the doped bismuthates is that short-range CDW correlations survive at high doping levels [3,5,6,7], and that these can be viewed as polarons – complex bound states of charge carriers and oxygen breathing distortions [8]. The metallic phase would then amount to a polaronic liquid, with the polarons binding into bipolarons that, in turn, condense to form a superfluid at $T_c$ [3,9]. On the other hand, it has been argued that superconductivity arises through a conventional electron-phonon pairing mechanism [10,11], from a homogeneous metallic state well-described by Fermi-liquid theory. Microscopically, the two pictures are radically different, yet conclusive experimental evidence for either scenario is lacking.

Here we use diffuse x-ray scattering and Monte Carlo modeling to study short-range structural correlations in $Ba_{1-x}K_xBiO_3$ (BKBO). We find that static and low-energy breathing distortions are absent in samples with a sufficiently high potassium (and, hence, hole) concentration, on either side of the insulator-metal boundary. This places strong constraints on any local CDW scenario and the associated polaronic effects. However, we uncover correlations of a different nature that involve anticorrelated bismuth and oxygen shifts and break inversion symmetry (Fig. 1c). We show that these distortions are the result of a complex energy balance that involves screened ionic interactions, precipitated by random potassium-barium substitution. The short-range structural symmetry is therefore considerably lower than the average symmetry, implying that BKBO is a locally non-centrosymmetric superconductor. This finding raises the possibility of unanticipated channels for electron-lattice interactions, with similarities to materials with hidden spin-orbit coupling [12,13] and systems without global inversion symmetry [14] or close to an inversion-breaking instability [15,16,17].

Figure 2 shows representative diffuse scattering data for two BKBO samples on either side of the metal-insulator transition. Qualitatively, the insulating samples show remarkably little diffuse scattering overall, indicative of nearly absent short-range correlations. In contrast, distinct features are immediately obvious in the metallic sample, with a notable asymmetry of the diffuse intensities around the Bragg peaks – the scattering is always stronger on the low-wavevector side of the peaks. This is a signature of distortions induced by local lattice contraction or expansion, known generically as size-effect scattering [18]. Yet in our case this is not simply due to Ba-K substitutional disorder, since the ionic radii of Ba and K are nearly the same. In fact, the $K^+$ ion is slightly *larger* than $Ba^{2+}$ while having a smaller x-ray form factor, which should lead to size-effect scattering that is *stronger* on the high-wavevector side [18]. Moreover, the asymmetric scattering shows a nontrivial doping dependence, with a strong increase in the metallic phase, and weak temperature dependence in the studied range (Fig. 2e). In contrast to Pb-doped $BaBiO_3$, which displays metastable structures and structural phase separation on the 10-20 nm scale [2,19], BKBO is thought to be single-phase in the metallic/superconducting state [2,20]. We note that the intensity maximum around $x \approx 45\%$ could be partially due to experimental uncertainties (see Methods), and that the $x \approx 38\%$ sample, which lies close to the insulator-metal boundary, might have a nonzero insulating volume fraction, which would decrease the integrated intensity.

Importantly, no diffuse signal is present around (½ ½ ½) and equivalent positions in any of the studied samples (Extended Data Fig. S1). This demonstrates, within our experimental sensitivity, the absence of short-range breathing distortions that double the quasi-cubic unit cell. Instead, we observe sharp, weak superstructure peaks that originate from a qualitatively different structural distortion – small long-range octahedral tilts – known from previous work [20]. Diffuse x-ray scattering is an energy-integrating probe and, in principle, sensitive to distortions at any energy scale. Yet the corresponding structure factors decay fairly quickly with the energy of lattice excitations [21], and dynamic distortions therefore become difficult to observe above 10-20 meV.

We therefore cannot rule out the presence of breathing distortions at higher energies, and breathing phonon modes are certainly present (between 70 and 80 meV [10]). However, the absence of low-energy distortions still places stringent constraints on polaronic theories of the metallic state, especially for small polarons with high binding energies and low mobilities, where significant distortions would be expected [8]. Moreover, our result for the insulating phase is certainly not consistent with a frozen polaronic state.

We can achieve a more refined understanding of the local structure by generating a three-dimensional vector pair distribution function (3D-ΔPDF) using established punch-and-fill methods to eliminate contributions from the average structure (see Methods). The 3D-ΔPDF therefore only shows short-range pair correlations that deviate from the average structure and enables model-free insight into the real-space local structure [22,23]. Two representative planes are shown in Fig. 2 for the $x \approx 45\%$ sample, with distinct pair correlations annotated. It is immediately clear that the characteristic length of the correlated displacements is short: the signals decay significantly within 2-3 unit-cell distances, *i.e.*, about 1-2 nm (see also Extended Data Fig. S2). Due to the simple crystal structure, it is straightforward to identify most of the features in the 3D-ΔPDF, especially at non-integer positions where there is no multiplicity. In the $z = 0$ plane, Ba-O correlation peaks are clearly visible, and the dipolar shape of the features shows that Ba and O atoms are more likely to be closer to each other than in the average structure. Note that K-O correlations would appear at the same positions, but the Ba-O signal dominates due to the much larger form factor of Ba. In contrast, Bi-Ba correlations (visible in the $z = ½$ plane) show that Bi and Ba are more likely to be further apart. Finally, Bi-O correlations do not involve appreciable changes of the interatomic distance, but rather a displacement of oxygen *perpendicular* to the average Bi-Bi line. This is seen as a four-fold symmetric pattern around (0 0 ½) and also quite clearly in the next-nearest-neighbour Bi-O correlations around (0 1 ½) and equivalent positions.

Two important conclusions can be drawn from the 3D-ΔPDF. First, features that would correspond to breathing distortions of the Bi-O bonds are absent, as already seen from the reciprocal space data (Extended Data Fig. S1). Second, the only way to reconcile the Bi-O correlations with the opposite signs of the Ba-O and Ba-Bi correlations is a displacement pattern that breaks inversion symmetry. In local environments where Ba and K are not symmetrically arranged, the inversion-breaking Bi-O distortion of Fig. 1c occurs, with the oxygen atoms shifting toward the Ba-rich side of the unit cell, and the Bi atoms shifting away from it (but with a much smaller amplitude). The Ba ions then adjust their positions to be further away from Bi, but the oxygen shift is larger than the Ba shift, so the Ba-O distance is still smaller than in the average structure. Importantly, a purely rigid $BiO_6$ octahedra tilts would not lead to pronounced Ba-O size-effect features (that are even visible, *e.g.*, in the next-nearest-neighbor Ba-O peaks around (1 ½ ½)), but yield a symmetric Mexican-hat feature centered at the origin in the 3D-ΔPDF [22]. Yet it is possible that a local rigid tilt component is superposed with the inversion-breaking displacement.

We obtain further microscopic insight through classical Monte Carlo modeling of the local structure, and first explore a toy model of (static) short-range CDW correlations using an effective Ising-type Hamiltonian based on Bi valence disproportionation. The bismuth valences +3 and +5 play the role of Ising spins, and K substitution disrupts the CDW, since it locally favors the +5 state. Free parameters of the model (see Methods) can be adjusted so that long-range CDW order disappears at low K concentrations, as expected, but short-range correlations survive up to higher K concentrations. We also introduce the accompanying breathing distortion by using a valence-dependent interaction and relaxing the Bi-O distances using Monte Carlo importance sampling. This invariably generates superstructure peaks at (½ ½ ½) and equivalent reciprocal space positions even for weak short-range correlations (Extended Data Fig. S1), *i.e.,* a signature of local CDW correlations that is clearly absent in our data.

In order to capture the observed diffuse scattering features, we then construct a minimal model that only includes Bi and Ba/K displacements, under the assumption that all Bi atoms are equivalent. Furthermore, it is assumed that local lattice distortions are caused by incomplete screening of the local ionic charges, which leads to electrostatic repulsion that is stronger for Bi-Ba pairs than for Bi-K pairs. This is achieved with a screened Coulomb (*i.e.*, Yukawa) potential to describe ionic interactions (see Methods). In the relaxed structure, Ba and Bi then prefer to be further apart than K and Bi (Fig. 3b). Since the scattering is dominated by the heavy atoms, the main reciprocal space features are well reproduced (Fig. 3a), with stronger diffuse scattering on the low-wavevector side of the Bragg peaks.

We further investigate the role of oxygen by extending the minimal model to include oxygen displacements perpendicular to the Bi-Bi direction. Using the bond valence sum method to estimate the oxidation states of Bi, Ba and K (see Methods), we find Bi and K oxidation states in reasonable agreement with their nominal values, whereas Ba is significantly underbonded. At low K concentrations, the system compensates for this through symmetry-lowering structural transitions [7], as permitted by the known size mismatch of Ba to the surrounding oxygen cage (Extended Data Fig. S3). On the other hand, at sufficiently high K concentrations, long-range tilt formation is disrupted by randomly distributed K atoms. Inversion-breaking oxygen displacements then become an alternative way to compensate for the underbonding of Ba in unit cells with an asymmetric arrangement of Ba and K. We include this in the Monte Carlo modeling by employing the squared bond valence mismatch to penalize chemically unstable coordination environments (see Methods). The obtained local structure contains a mixture of rigid tilts and inversion-breaking displacements that decrease the Ba-O distances (Fig. 3c), in agreement with the 3D-ΔPDF analysis. The inversion-breaking distortions are found to increase with increasing K concentration (Extended Data Fig. S5). We emphasize that the random K/Ba substitution and short-range electrostatic interactions are crucial in generating the distortions.

Our results establish that bismuth valence disproportionation and oxygen breathing-mode polarons are likely not relevant to superconducting BKBO. However, other electron-phonon composite states could be present, particularly states that involve the inversion-breaking Bi-O phonon mode. *Ab initio* studies suggest that the coupling of electrons to the longitudinal Bi-O mode around 20 meV is in fact stronger than the coupling to high-energy oxygen breathing modes [10], which may lead to polaronic physics. The appearance of a broad infrared peak in the optical conductivity [24] is consistent with electron-phonon bound states. However, it is also possible that the local inversion-breaking distortions are predominantly precipitated by structural properties, with electron-phonon coupling playing a secondary role and the metallic state more conventional. This is supported by the increase of inversion-breaking correlations with doping in our simulations, which do not explicitly include electron-phonon coupling. A related, important question is the interplay between the inversion-breaking displacements and rigid octahedral tilts – the latter are caused by atomic-size mismatch and decrease with doping, whereas the former clearly increase substantially in the metallic phase (Fig. 2e). It is therefore possible that the rigid tilts compete with the polar displacements, which might then only proliferate below a critical tilt value, with a direct relation to the insulator-metal transition. An analogous interplay between octahedral tilts and electronic properties has been observed in, *e.g.*, certain cuprates [25]. Targeted uniaxial-strain experiments to influence the tilts, as recently demonstrated in rare-earth titanates [26], might provide more information about this aspect of bismuthate physics, especially given the possible role of tilts in enhancing electron-phonon coupling [27]. Another interesting possibility is that screening by mobile carriers is essential to obtain the local distortions. Our calculations use the screened Yukawa potential for electrostatic interactions, which severely limits their spatial range and permits inhomogeneity on the scale of the screening length. This is appropriate in the metallic phase, but not in the insulating phase, where long-range Coulomb interactions likely favour configurations with significantly less short-range charge disorder, which might explain the enhancement of diffuse scattering for metallic compositions.

The existence of local correlated inhomogeneity has profound implications for both the metallic and the superconducting states. Similar to other prominent metallic perovskites, such as titanates [28], nickelates [29], and vanadates [30], the resistivity of the bismuthates is approximately proportional to the temperature squared, consistent with Fermi-liquid physics, but with a magnitude that is unrealistically large for electron-electron scattering [31]. It has been proposed that this might generically be understood through the interaction of small polarons with low-lying phonon modes [30]. The phonon spectrum of the bismuthates contains several optical modes in the 10-20 meV range [10] that could lead to such behaviour, but the polarons would probably involve the inversion-breaking displacements instead of breathing distortions. Moreover, local inversion-symmetry breaking can cause electron-lattice interactions that are forbidden in the average high-symmetry structure; electron-phonon coupling in a conventional Fermi liquid can also enhance $T^2$ resistivity [32]. Importantly, the Thomas-Fermi screening length in metallic BKBO is about 4 Å [7] and likely sets the structural correlation length, implying that the dipole moment associated with the inversion-breaking distortions is not completely screened. A Rashba-type electron-phonon coupling that is linear in phonon amplitudes might be important for both normal-state properties and superconducting pairing, as proposed for materials close to a polar instability such as strontium titanate [15,33]. Moreover, since the inversion-symmetry breaking appears at length scales comparable to the superconducting coherence length of 30-60 Å [2], the bismuthates can be viewed as locally non-centrosymmetric superconductors. This indicates that the pairing might be more exotic than just simple *s*-wave, with admixtures of odd-momentum gap functions allowed by symmetry [14], and the appearance of local time-reversal symmetry breaking. Finally, it will be interesting to see if similar effects are present in $BaPb_{1-x}Bi_xO_3$ and the newly discovered isostructural antimony-based superconductor $(Ba,K)SbO_3$ [34], as well as in substitutionally-doped oxides more broadly.

**Methods**

*Samples.* Ba$_{1-x}$K$_x$BiO$_3$ crystals were grown using a low-temperature electrocrystallization method similar to previous work [35,36]. The growth precursor consisted of 140 g KOH, which acts as flux, 10.5 g Bi$_2$O$_3$, and 3.5-20 g Ba(OH)$_2 \cdot$ 8H$_2$O. The precursor materials were combined in a PTFE crucible, melted at 400 °C and homogenized by a magnetic stirrer at 60 rpm in a humid nitrogen atmosphere. A small amount (2 - 4 g) of KNO$_3$ was added to counteract the degradation of the crucible [36]. Platinum electrodes were inserted into the molten solution and a constant current between 0.1-1 mA was supplied for 18 hours. BKBO crystals deposit on the anode, and excess bismuth deposits on the cathode. The amount of Bi$_2$O$_3$ far exceeds the solubility limit of KOH at the beginning of a growth, enabling the Bi$_2$O$_3$ concentration to remain constant for the duration of the growth. The ratio of Ba(OH)$_2 \cdot$ 8H$_2$O to KOH controls the doping level of the deposited crystal. Using this method, crystals with potassium content between $x = 0.25$ and $0.51$ were grown.

Sharp superconducting transitions from dc magnetization and ac susceptibility (including nonlinear susceptibility [37]) measurements were used as an indicator to select metallic samples. For both metallic and insulating samples, high crystalline quality was established from sharp, well-defined Bragg peaks in the x-ray data. The potassium concentration was estimated from lattice parameter values obtained during the refinement of the diffuse x-ray scattering data, and by employing the known linear relation between doping and unit cell size at 300 K [7]. In the case of metallic samples, these values were found to be consistent with those determined from the well-defined relationship between $T_c$ and doping [7].

*Diffuse x-ray scattering.* The measurements were performed on beamline 6-ID-D of the Advanced Photon Source, Argonne National Laboratory, USA. The beamline uses a superconducting magnet undulator insertion device and a double-crystal monochromator to produce a monochromatic high-energy x-ray beam. We used a photon energy of 87 keV, slightly below the Bi K-edge at 90.526 keV, which likely suppresses the relative intensities of the diffuse scattering/PDF peaks that involve Bi. A Pilatus 2M CdTe detector with sensor layer optimized for high energies was used to collect frames with 0.1 s exposure time, while samples were continuously rotated at 1° s$^{-1}$ about a horizontal axis. Three sets of rotations were performed in each measurement, between which (*i*) the detector was translated by 5 mm in both the horizontal and vertical directions to cover gaps between detector chips, and (*ii*) the sample rotation axis was offset by ±15° from perpendicular to the beam to allow masking of artefacts caused by scattering in the sensor layer. The overall counting time at each temperature setting was 20 min. See Supplementary Material in [23] for more details. Data covering a range of about ± 15 Å$^{-1}$ in all directions were collected down to 30 K using an Oxford N-Helix helium cryocooler, and transformed to S(**Q**) using the software package CCTW (https://sourceforge.net/projects/cctw/).

*3D-ΔPDF.* The 3D-ΔPDF shown in Fig. 2 was generated similar to [23] by symmetrizing the $x = 45\%$ data with all operations included in the $m\bar{3}m$ point group, "punching" out spheres of reciprocal space of radius 0.2 r.l.u. around each Bragg position, and "filling" the removed spheres by interpolating with a three-dimensional Gaussian kernel. A Tukey window (α = 0.5) was applied to the "filled" data to reduce leakage artifacts in the Fourier transform, and a discrete Fourier transform was then applied to the data. The 3D-ΔPDF shown is the real part of this Fourier transform.

The real-space correlation length of the displacement was obtained from a representative linecut ($0.5 \leq x \leq 6$, $y = 0$, and $z = 0$) of the 3D-ΔPDF. The absolute values of the signal peaks along the linecut were fit to an exponential function (Extended Data Fig. S2). The decay constant was found to be 1.1 unit cells, which indicates a significant decay within about 2 unit cells (about 1 nm).

*Temperature and doping dependences of diffuse scattering.* The temperature and doping dependence of the asymmetric component of the diffuse signal was obtained as follows. First, a Bragg peak free of any artefacts such as streaking (due to Compton scattering) or powder rings was chosen. For a comparison of relative scattering intensities among different samples, it is essential that the same Bragg peak and Brillouin zone be used, as there exists considerable intensity variation across different zones. Because BKBO is nearly cubic at the measured temperatures, there are 8 equivalent sites for each *hkl* position in reciprocal space, and the positions ($\pm 4\ 0 \pm 6$) were chosen. A one-dimensional (1D) intensity profile along [100] was obtained by summing between $\pm\ 0.04$ r.l.u. along [010] and $\pm\ 0.08$ r.l.u. along [001]. The result in Fig. 2(e) was obtained from fast-Fourier transformations of these 1D datasets, centered on the respective Bragg peak, *via*

$$y(k) = \sum_{n=0}^{N-1} e^{-2\pi i \frac{kn}{N}} x(n), \tag{1}$$

where *N* is the length of the 1D array, *x* the intensity at index *n*, and $0 \leq k \leq 41$ (this is well into the range of *k* where the contribution of each higher value becomes vanishingly small). This transformation returns both real values, which correspond to the amplitudes of the symmetric parts of the signal, and imaginary values, which correspond to the asymmetric scattering intensities. For each sample at each temperature, the asymmetric intensities were summed and normalized by the maximum value in the 1D dataset to obtain a 'temperature factor'. This normalization functioned quite well for a comparison of datasets for the same sample; however, a second normalization was required to compare results for different samples to account for differences in illuminated sample volumes and background levels. This normalization consisted of finding the Bragg peak intensity at the same temperature for each sample and scaling each sample's temperature factors by the ratio of its Bragg peak intensity to that of the sample with the strongest Bragg peak intensity. Notably, this procedure involved uncertainties that were difficult to quantify, most importantly possible partial saturation (nonlinearity) of the detector at the Bragg peaks for the largest samples, which might affect the comparison in Fig. 2e. The relative intensities should therefore be taken as semi-quantitative, with the main, robust result that the diffuse intensity increases in the metallic phase.

*Monte Carlo modeling.* We used classical Monte Carlo (MC) to simulate short-range ordered structures [38], with K randomly substituted for Ba. As a first step, a clear diffuse scattering signature of short-range CDW correlations was established by constructing a simple, yet robust Ising-type model with the Hamiltonian

$$H = J_1 \sum_{\langle ij \rangle} s_i s_j + J_2 \sum_{\langle ij \rangle} \sigma_i s_j, \qquad (2)$$

where bismuth oxidation states +3 and +5 are mapped to the respective values $+1$ and $-1$ of the spin variable $s$, the chemical species of nearest-neighbor A-site cations are encoded as $\sigma=+1$ (K) and 0 (Ba), the indices $i$ and $j$ label different unit cells, and the sums are restricted to the nearest-neighbor pairs. Since we aim for a phenomenological model of short-range CDW formation, the coupling constants $J_{1,2}$ can be estimated from the requirement that long-range order disappears for K concentrations above ~0.1 [15,20], while short-range CDW correlations persist at higher doping. Long-range CDW order corresponds to a nonzero expectation value of the staggered magnetization

$$m = \frac{1}{N} \langle \sum_{ijk} (-1)^{i+j+k} s_{ijk} \rangle, \qquad (3)$$

where $N$ is the total number of unit cells, and the brackets denote the ensemble obtained via MC importance sampling. Similarly, short-range order is captured by the chemical correlation coefficient [39]

$$c_{ij} = \frac{P_{ij} - \theta^2}{\theta(1-\theta)} \qquad (4)$$

where $P_{ij}$ is the joint probability that bismuth sites $i$ and $j$ are occupied by the same oxidation state +3 (+5), and θ is the overall concentration of +3 (+5) atoms. Negative values of $c_{ij}$ correspond to situations where the two sites tend to be occupied by different oxidation states, which is used to quantify remnant CDW correlations. For the sake of computational efficiency, the model neglects the possibility of partial charge transfer between different $BiO_6$ octahedra and operates only with discrete values of charge perfectly localized on Bi atoms. Though likely unrealistic, this assumption still serves our purpose, since the relevant diffuse features will mostly be determined by the range of CDW correlations and the octahedral breathing amplitude, and not by the exact Bi oxidation states. The charge should thus be viewed as an effective charge variable. The first term in the Hamiltonian ($J_1 > 0$) models the Coulomb interaction between nearest-neighbor charges localized on Bi atoms and is thus responsible for CDW formation. The second term ($J_2 < 0$) captures the disruption of the CDW order by the presence of K atoms, which play the role of effectively negatively charged impurities and locally favor the +5 state. A representative statistical ensemble of equilibrium configurations was obtained by simulating many $20 \times 20 \times 20$ supercells with periodic boundary conditions, such that the basic move involves a swap of +3 and +5 atoms in order to preserve the effective charge balance, and by using simulated annealing [40] to prevent the simulation from getting stuck in local minima. Furthermore, we used the coupling between the local charge and octahedral breathing amplitude [41,42] that is responsible for stabilizing the charge modulation, through the valence-dependent harmonic interaction between Bi and O atoms,

$$U_i(r) = \frac{1}{2}K[r - r_0(s_i)]^2, \tag{5}$$

where $K$ is the effective spring constant and $r_0(s_i)$ the Bi-O equilibrium distance for two different oxidation states $i$. The value of the spring constant $K \approx 19$ eV/Å$^2$ can be estimated by fitting the Raman-active 70 meV breathing mode [42,43], and the average values of $r_0(+5)$ and $r_0(+3)$ in the parent compound are known from previous studies [2,7,20] to be ~ 2.11 Å and ~ 2.29 Å, respectively. Since we aimed to reproduce the local structure that would result from short-range CDW, the above valence-dependent potential was sufficient to reproduce the relevant effects and avoided microscopic details of electron-phonon coupling and the intricacies of stability of polaronic and bipolaronic solutions [3,5]. Using MC sampling with positional displacements as the basic move to relax the Bi-O bonds, and subsequently calculating diffuse scattering [44], we found diffuse superstructure peaks at specific half-integer positions in reciprocal space, which are absent in the experiment (Extended Data Fig. S1). The configuration and diffuse scattering shown in Extended Data Fig. S1 were obtained using $|J_2/J_1| = 4$ and $J_1$ equal to the effective Monte Carlo temperature. For this choice of parameters, no long-range order is present, but short-range correlations persist (Extended Data Fig. S1). The results do not significantly depend on the ratio of $J_1$ to the temperature, if the ratio is smaller than or close to one.

The second part of the analysis involved a minimal model that captures the essential features of the observed diffuse scattering (Fig. 2). First, since we find no evidence of charge modulation in the studied K concentration range, we assumed a homogenous charge distribution on the Bi sublattice, with the oxidation state of Bi fixed by charge neutrality ($V_{Bi} = 4 + x$). The pseudo-cubic lattice constant obtained from this assignment using bond valence sum rules [45] for the BiO$_6$ octahedral complex is consistent with $a = 4.35 - 0.17x$ known from structural studies [7], which indicates that the unit cell size is mainly controlled by the strength of the Bi-O bonds and gives credibility to our assumption. Also, the asymmetry of the observed diffuse features indicates that the local lattice distortion is not simply due to slight differences in the ionic sizes of Ba$^{2+}$ (~ 1.49 Å) and K$^+$ (~ 1.52 Å) (Shannon-Prewitt ionic radii) [46]. Therefore, we assumed that the effect is caused by residual electrostatic interactions between different non-bonded ions, where the static screening in the metallic regime is approximately accounted for by the familiar Yukawa potential

$$\phi_{ij}(r) = \frac{Q_i Q_j}{r} e^{-k_S r}, \tag{6}$$

where $Q_i$ represents the ionic charge of the i-th ion and $k_S \approx 0.49$ Å$^{-1}$ [20] the screening wavenumber, which sets the length scale of the local lattice distortion. In the insulating state, the screened potential is likely not appropriate, and long-range Coulomb interactions might drastically change the overall energy balance; this is, however, significantly more difficult to simulate, and was not been attempted. The ionic charges were assumed to be equal to atomic oxidation states, which suffices for the desired semi-quantitative insights that we aimed for. To make the model more realistic, we introduced Bi-O, Ba-O and K-O covalent bonds via the Morse pair potential

$$U(r) = D\left(1 - e^{-(r-r_0)/b}\right)^2, \tag{7}$$

where the equilibrium distance between atoms $r_0$ is determined by the size of the unit cell. The constant $b$ determines the "softness" of the bond and was taken from [47]. $D$ was obtained by comparing to the known values of atomic displacement parameters (thermal factors) [7,20]; $D_{Bi-O} = 10$ and $D_{Ba-O} = D_{K-O} = 2$ in units of MC temperature. The structure was relaxed by performing the MC thermalization on an ensemble of $15 \times 15 \times 15$ supercells. We found that the calculated diffuse scattering averaged over many equilibrium configurations and symmetrized in the m3m point group agreed well with the data, with the asymmetry on the low-wavevector side of the Bragg peaks reproduced well (Fig. 3a).

To complete the picture, we constructed a model of the local structure of oxygen octahedra, whose features are not obvious in reciprocal-space data, but can be inferred from the 3D-ΔPDF. We used the well-known perovskite ($ABO_3$) tolerance factor

$$t = \frac{r_A + r_O}{\sqrt{2}(r_B + r_O)} \tag{8}$$

as an indicator of structural stability, where $r_O = 1.26$ Å [46] is the ionic radius of $O^{2-}$. For $r_B + r_O$, we used half the lattice constant, since the size of the unit cell is controlled by the covalent Bi-O bond whose length depends on doping through the effective valence of Bi. Using tabulated values of ionic radii of $K^+$ and $Ba^{2+}$ in place of $r_A$ we obtained the doping dependences of $t_{Ba}$ and $t_K$ (Extended Data Fig. S3), and hence an indication of how well the A-site cations are accommodated by the lattice. $t_K$ is in the "ideal" range for all K concentrations, whereas Ba is too small for the ideal cubic lattice, which can be compensated for by the formation of long-range rigid octahedral tilts that lead to the sequence of structural transitions ($Pm3m \rightarrow I4|mcm \rightarrow Ibmm \rightarrow I2_1m$) [7,20] as the K concentration decreases toward zero. However, for higher K concentrations, formation of long-range ordered tilts is significantly disrupted by the presence of randomly distributed K atoms. The barium size mismatch and the disorder introduced by potassium doping thus lead to an interplay between short-range tilts and inversion-breaking displacements. This is also reflected in the unusually large anisotropic thermal ellipsoids of oxygen [7], indicating the propensity for large oxygen displacements perpendicular to the Bi-O bond to accommodate the size discrepancy. Similar conclusions can be drawn if we use the bond valence sum (BVS) method [45,48] to estimate the empirical valence $v$ of a given atom as a sum of contributions of each bond that it forms (bond valences $s_j$) in each bonding environment, with $v = \sum_j s_j$ the scalar and $\mathbf{v} = \sum_j s_j \mathbf{e}_j$ the vector bond valence sum, where $\mathbf{e}_j$ are unit vectors along the bonds, and the summation is over bonds. The bond valences are best understood in the context of (mostly) ionic bonds where bond valence plays the role of electric flux, so that the BVS rule above simply becomes Gauss' law. Using the known correlation between bond valence and bond length

$$s_j = e^{\frac{R_0 - r}{b}}, \tag{9}$$

where $r$ is the bond length and $R_0$ and $b$ are parameters tabulated in [47], we calculated the doping dependence of the empirical Ba and K valences in the average structure. In the doping range of interest ($x \sim 0.35 - 0.45$), we found K to be fully bonded ($v_K$ close to nominal valence +1 of K), and Ba to be significantly underbonded ($v_{Ba}$ smaller than nominal valence +2 of Ba) (Extended Data Fig. S3). This correlates with our previous geometric analysis, but also allowed us to formulate a semi-quantitative model where we use the (squared) valence mismatch, also referred to as the global instability index, defined by [45]

$$g = w_s \sum_i (v_i - V_i)^2 + w_v \sum_i (\mathbf{v}_i^2 - \mathbf{V}_i^2)^2 \tag{10}$$

as a measure of structural instability [48,49]. The two right-hand side terms correspond to scalar and vector contributions, respectively [49], with weights $w_s = 2$ and $w_s = 1$. $V_i$ and $v_i$ are the scalar nominal and empirical valences of an atom, respectively, while $\mathbf{V}_i$ and $\mathbf{v}_i$ are the respective vectors. The nominal vector valences $\mathbf{V}_i$ are taken to be zero [45,49]. Using $g$ in the calculations naturally captures multi-body atomic correlations and the tendency of oxygen displacements to compensate for the valence mismatch of Ba. To constrain the model, we implemented soft volume exclusion using the Lennard-Jones repulsive term $A(r_{ion}/r)^{12}$, where $r_{ion}$ is the tabulated ionic radius of a given atom [46], and $A = 6$ in units of MC temperature. Furthermore, the interplay of tilts and polar displacements required the inclusion of an effective octahedral rigidity constraint, for which we used a harmonic spring potential between neighboring oxygen atoms as the simplest option,

$$U_R(r) = \frac{1}{2}k(r - r_O)^2, \tag{11}$$

where $r$ is the O-O distance, $r_{eq}$ is their distance in the average structure, and the spring constant $k = 10$ (in MC temperature units) was chosen to be sufficiently large, thus penalizing, but not precluding non-rigid displacement patterns of the oxygen octahedra. Given the qualitative nature of the model, the exact proportion of rigid tilts as opposed to non-rigid polar distortions is hard to estimate, as it heavily depends on the relative weights of $g$ and $U_R(r)$. Still, overall trends in the doping dependence can be discerned either by looking at the distribution of O-O distances (Extended Data Fig. S4) or the doping dependence of the local centrosymmetry parameter, defined as [50]

$$p_{CSP} = \sum_1^{N/2} |\mathbf{r}_i + \mathbf{r}_{i+N/2}|^2 \tag{12}$$

where $N$ is the number of nearest neighbours of a given atom, and $\mathbf{r}$ are vectors from this central atom to a pair of opposite neighbours. This parameter is a convenient measure of deviations from centrosymmetry, and it increases with K concentration (Extended Data Fig. S5).

**Acknowledgements**

We thank D. Robinson and S. Rosenkranz for assistance with x-ray scattering experiments, and T. Birol, A. Klein, and S. Johnston for discussions and comments. The work at the University of Minnesota was funded by the U.S. Department of Energy through the University of Minnesota Center for Quantum Materials, under Grant No. DE-SC0016371. The work in Zagreb was supported by Croatian Science Foundation Grant No. UIP-2020-02-9494. The work at Argonne was supported by the U.S. Department of Energy, Office of Science, Basic Energy Sciences, Materials Sciences and Engineering Division. This research used resources of the Advanced Photon Source, a U.S. Department of Energy (DOE) Office of Science User Facility operated for the DOE Office of Science by Argonne National Laboratory under Contract No. DE-AC02-06CH11357.

**Author Contributions**

MG and DP conceived the work. SG, JJ, DZ and DP grew and characterized samples. SG, JJ, ZA, MJK, DP and RO performed x-ray scattering experiments and analysed data. MS performed model calculations. DP, MG, SG and MS wrote the manuscript with input from all authors.

**Competing Interests**

The authors declare no competing financial or non-financial interest.

**Data availability**

All data, materials and computer code used to generate the results in the paper are available from the corresponding authors upon request.

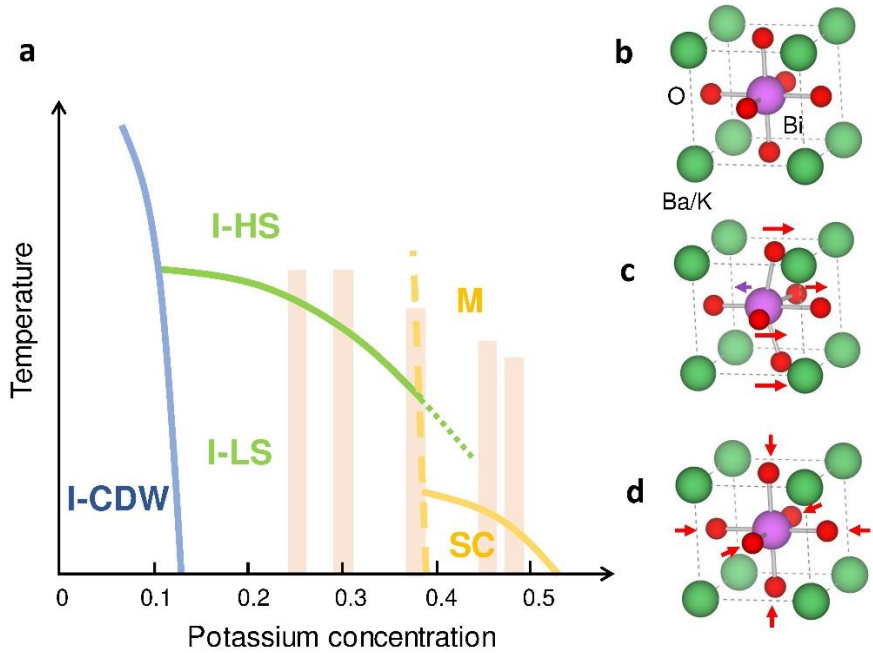

**Fig. 1 | Phase diagram and unit-cell distortions in $Ba_{1-x}K_xBiO_3$. a,** Schematic phase diagram based on [2,7,20]. In the insulating (I) phase, there exists a commensurate charge-density wave (I-CDW) up to $x \sim$ 10-15%. Metallic (M) and superconducting (SC) phases appear above $x \sim$ 35%. At intermediate K concentrations, long-range CDW correlations are absent [7]; the high-temperature structure (I-HS) is cubic, whereas long-range rigid octahedral tilts lead to tetragonal and orthorhombic distortions at low temperatures (I-LS; see Methods and [7,20] for details). The long-range tilts persist in the metallic phase (green dotted line). Shaded bands indicate estimated potassium concentrations for the samples studied in this work. **b,** Cubic unit cell. **c,** Schematic inversion-breaking distortion; two of the six oxygen atoms do not change position. **d,** Schematic oxygen-breathing distortion.

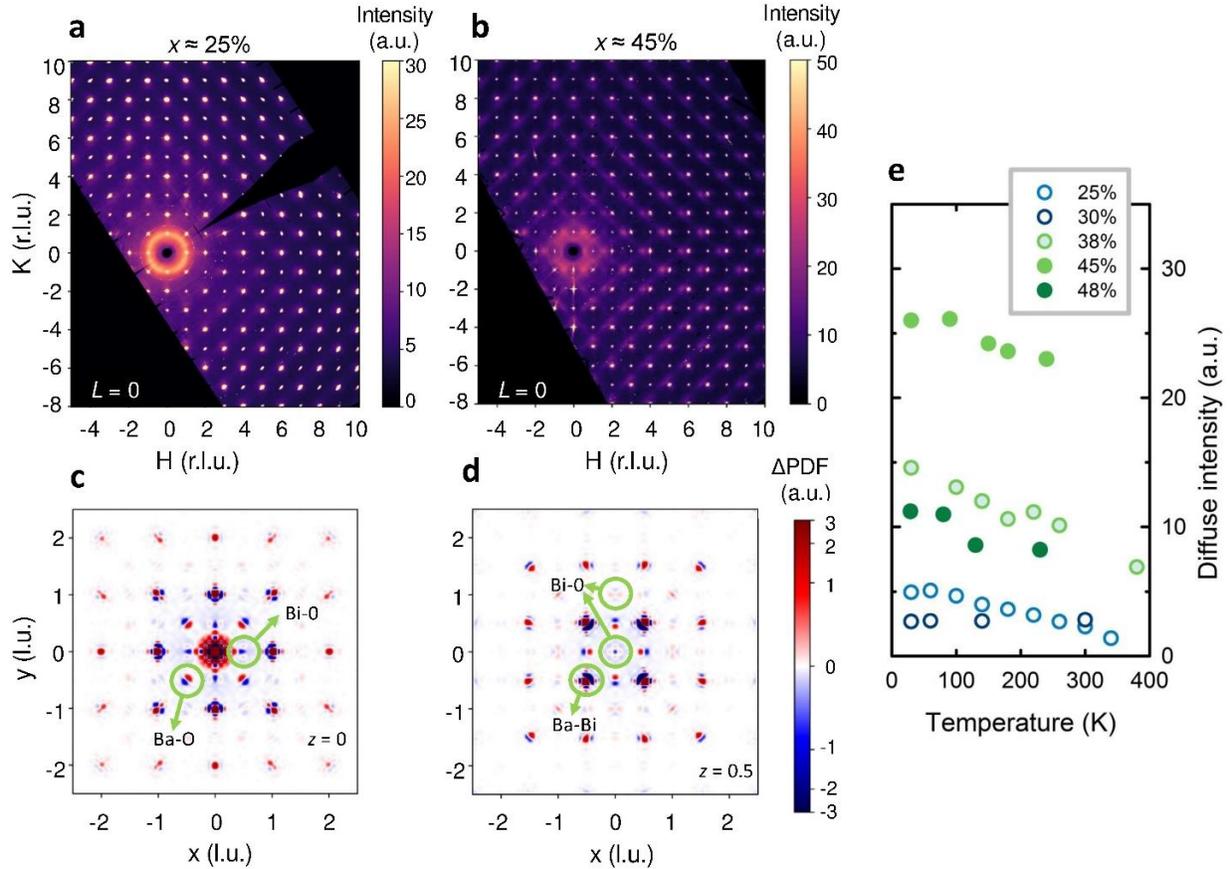

**Fig. 2 | Local structure of $Ba_{1-x}K_xBiO_3$. a,b,** Normalized diffuse x-ray intensity in the $L = 0$ plane at 30 K for (**a**) $x \approx 0.25$, in the insulating state, and (**b**) $x \approx 0.45$, in the metallic state. Structured diffuse scattering that is asymmetric around the Bragg peaks is clearly seen in the metallic sample. **c,d,** Vector pair distribution functions (3D-ΔPDF) obtained for the sample with $x \approx 0.45$ (same as in (**b**)), shown in two representative planes ($z = 0$ in **c** and $z = 0.5$ in **d**), with a number of Ba-O, Ba-Bi and Bi-O correlation peaks annotated. Correlations involving K are suppressed compared to Ba because of the significantly smaller x-ray form factor of K. **e,** Temperature dependence of the asymmetric component of the diffuse scattering across the insulator-metal boundary. Open and full symbols indicate insulating and metallic compositions, respectively; the $x \approx 38\%$ sample is close to the phase boundary. The intensities in (**a,b,e**) are normalized to nearby Bragg peaks; although it is difficult to extract absolute intensities from x-ray data, a strong increase in relative intensity is clearly seen in the metallic phase (see Methods).

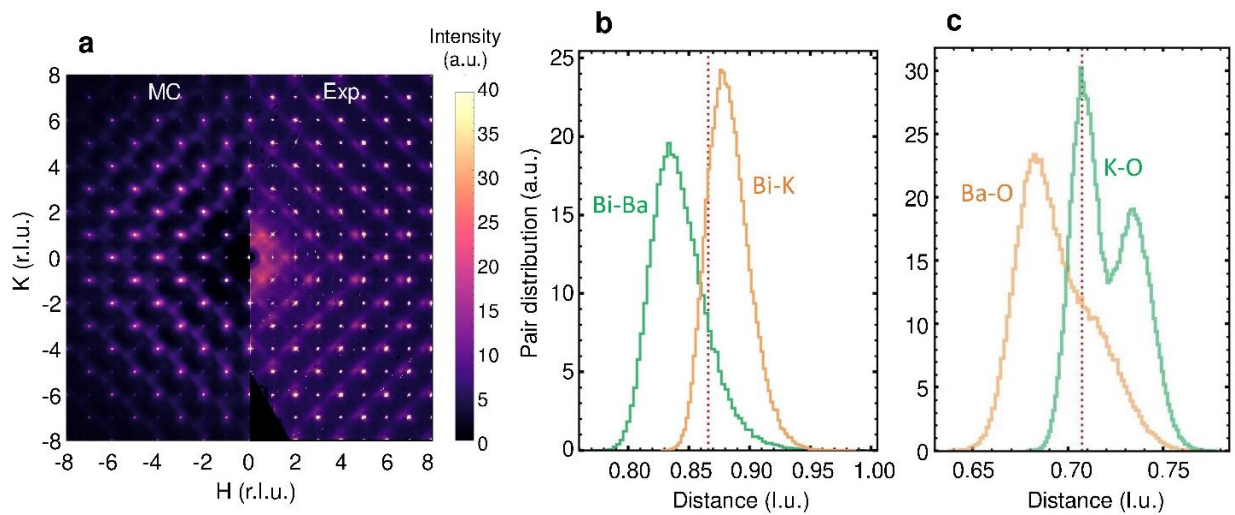

**Fig. 3 | Monte Carlo modeling of the local structure. a,** Calculated reciprocal-space image (left half) from a simple model (see Methods) that only includes negative Bi-Ba and positive Bi-K correlations. The model captures the main features observed in the experiment, in particular the diagonal rods of scattering and the asymmetry of the diffuse intensity around Bragg peaks (right half). **b,** Distribution of Bi-Ba and Bi-K distances within the model, showing a significant difference, which is the microscopic origin of the observed asymmetric diffuse features. Vertical dashed lines indicate the distances in the average structure. **c,** Distribution of nearest-neighbour Ba-O and K-O distances in an expanded model that includes oxygen displacements (see Methods). Due to inversion-breaking distortions, the Ba-O distances are shorter than average, in agreement with the experimental 3D-ΔPDF result. The calculated O-O distance distribution, which is not directly observable in the experiment due to small oxygen x-ray form factors, is shown in Extended Data Fig. S4.

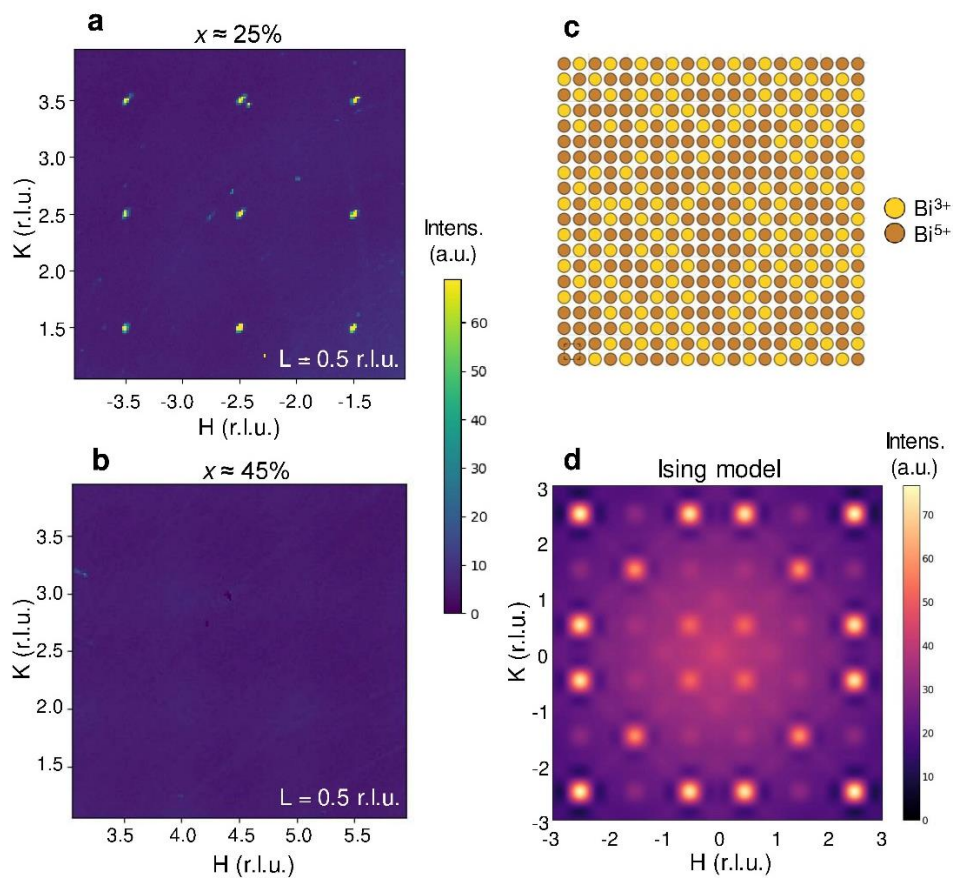

**Extended Data Fig. S1 | Short-range charge-density-wave features. a,b,** X-ray scattering data for BKBO with doping levels $x \approx 25\%$ and $45\%$, respectively, in the $L = ½$ plane. If short-range CDW correlations were present, broad diffuse peaks at half-integer positions would be observed. Such features are absent in both insulating (**a**) and metallic (**b**) samples. Instead, sharp superstructure peaks are seen in (**a**) due to a known long-range tilt distortion that is unrelated to the CDW and becomes very weak in the metallic/superconducting doping range [20]. The superstructure peaks appear at (½ ½ ½ ) and all equivalent positions due to the presence of tetragonal domains in the tilted phase; the tilt distortion in a single domain intrinsically does not double the lattice parameter, in contrast to the CDW. **c,** Typical equilibrium distribution of effective $Ba^{3+}$ and $Ba^{5+}$ ions in a simple Bi valence Ising model with frustration (one Monte Carlo relaxed configuration – see Methods for details). The K concentration is 30%. **d,** Diffuse scattering for the same plane as in **a** and **b**, calculated from Monte Carlo simulations of the Ising model with included oxygen breathing distortions, with effective K concentration of 35%. The model contains short-range CDW correlations which clearly lead to diffuse features, in stark contrast to the measured diffuse scattering.

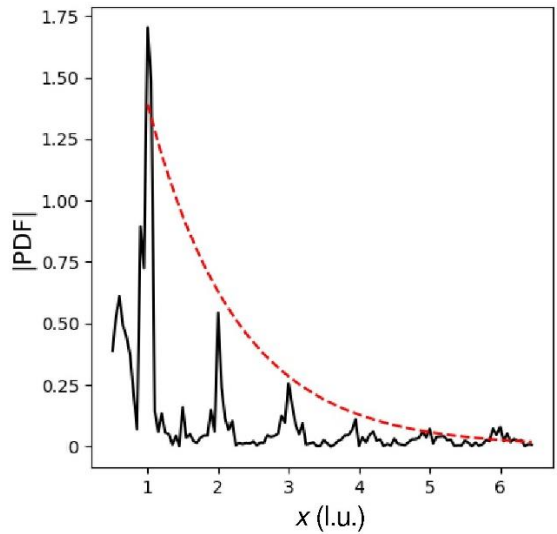

**Extended Data Fig. S2 | Structural correlation length.** The characteristic length for short-range correlations is obtained by fitting a simple exponential decay to the absolute value of the 3D-ΔPDF in Fig. 2 (sample with $x \sim 0.45$, 30 K). The representative cut shown here is $y/a = 0$, $z/a = 0$, and the decay constant from the fit is $1.1 \pm 0.1$ (error from fit), implying that the signal decays significantly within 2-3 unit cells (about 1 nm).

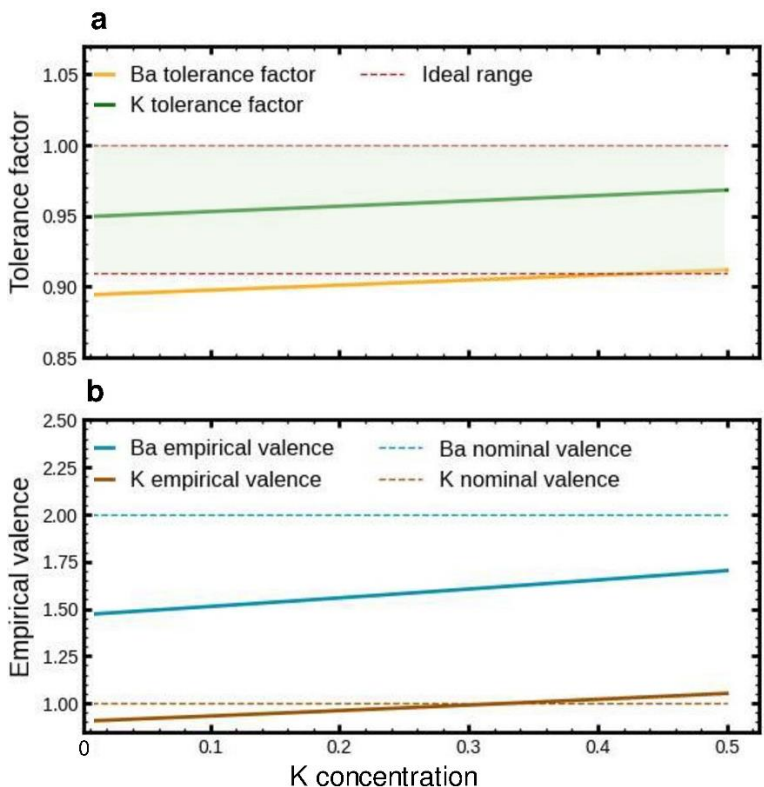

**Extended Data Fig. S3 | Structural and chemical parameters of Ba and K in the local-structure model**. **a,** Perovskite tolerance factor $t$ (see Methods), showing that the K ion size matches the ideal perovskite structure ($t = 1$) better than Ba, with further improvement upon doping. The shaded 'ideal range' is the approximate range where the cubic perovskite structure can be stabilized. The Ba size mismatch predominantly leads to rigid tilt distortions, which decrease with increasing K concentration. **b,** Empirical valence of Ba and K (see Methods), demonstrating that K is very close to the nominal valence of $+1$, whereas Ba is quite strongly underbonded. This leads to an effective attraction between Ba and O, which results in inversion-breaking displacements of the oxygen atoms in local environments with an asymmetric arrangement of Ba and K.

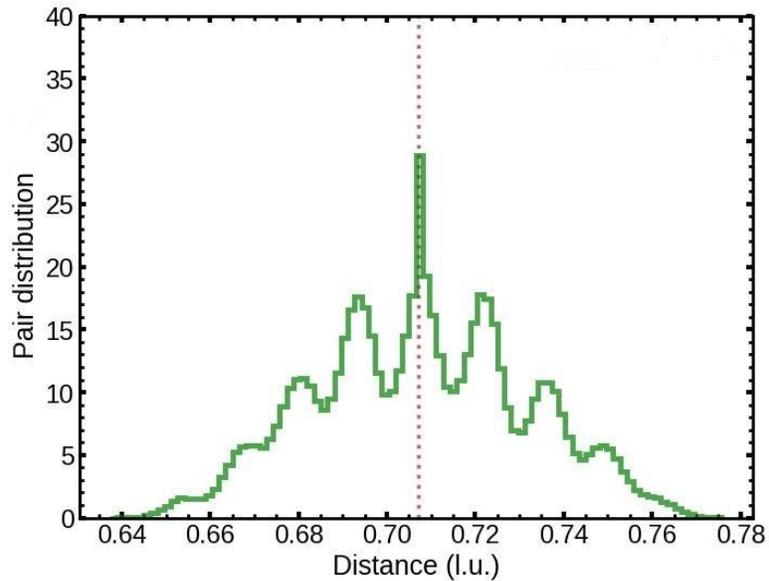

**Extended Data Fig. S4 | Calculated oxygen-oxygen correlations.** Distances between nearest-neighbour pairs of oxygen atoms obtained from Monte Carlo simulations (see Methods for details), similar to Fig. 3b,c. The vertical dashed line indicates the distance in the average structure. The individual peaks roughly correspond to inversion-breaking distortions in different local environments, *e.g.*, with different local Ba/K ratios and spatial configurations.

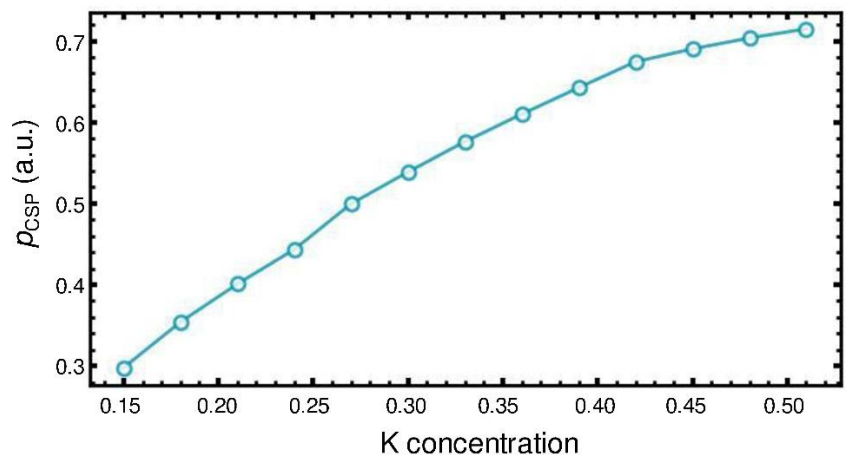

**Extended Data Fig. S5 | Calculated K concentration dependence of the inversion-breaking distortion.** The centrosymmetry parameter (see Methods), used as a direct measure of the inversion-breaking displacements, shows a substantial increase with K concentration, in qualitative agreement with the experimental results. The modeling does not take the insulator-metal transition into account, *i.e.*, we use screened electrostatic interactions appropriate for the metallic phase throughout.